# Open data adoption in Australian government agencies: an exploratory study

**Mohammad Alamgir Hossain**
School of Business IT and Logistics
RMIT University
Melbourne, Australia
Email: mohammad.hossain@rmit.edu.au

**Caroline Chan**
School of Business IT and Logistics
RMIT University
Melbourne, Australia
Email: caroline.chan@rmit.edu.au

## Abstract

Australia is among the leading countries that envisaged releasing unclassified public data under open license and reusable format with no further restriction on re/use. But, according to the Australian Information Commissioner John McMillan, Australia's progress on open data is 'patchy' and 'transitional'. He also evidenced that although a few agencies are proactive and have embraced the movements quite seriously, still there are "many obstacles that worked against effort to make government information and data discoverable and usable" (Hilvert 2013). Despondently, there is little empirical evidence that could explain what makes public departments not to release public data. Driven by the nature of the research, this study conducted an exploratory field study in Australia by interviewing eleven employees from six different government agencies. Applying content analysis technique, this study identifies six important antecedents to adoption of open data in public organisations, and proposes future research to test their relationships. As the main theoretical contribution, this study extends organisational behaviour toward technology diffusion. The findings of this study incite policymakers and managers to think about and prepare future strategies on open data developments.

## 1 Introduction

Citizens now are more concerned to democracy. They increasingly demand ownership to policymaking, and hence want access to public data (Zuiderwijk et al. 2014a). In fact, tremendous technological development and unprecedented explosion of peoples' computing skill in recent times, in terms of accessing, storing, manipulating, analysing, linking and distributing data, have been observed (Boulton 2014; Boulton et al. 2011; Rohunen et al. 2014) to made such access a reality. Accessing or using data is now comparatively simpler and easier than before with the explosive growth of mobile networks, and the resultant rise of social networks (Huijboom and Van den Broek 2011). Many applications developed or being developed for various electronic devices e.g. mobile, require access to various public data such as crime or accident stats, traffic data (Rohunen et al. 2014), train schedule, weather and environmental data (Mazumder 2014); on facilities including parks, toilets, locations of toxic waste dumps, public healthcare (Hendler et al. 2012); maps, satellite photographs, geographical locations, public sector budgeting, livestock and food-safety information, and so forth (Hendler et al. 2012; Janssen et al. 2012). Therefore, there are expectations that these data can be made publicly available for free and in reusable formats. Consequently, "the custodian of the public's data" do not have any choice but "are obliged to provide it" (MacGunigal 2014). Realising the benefits of providing data that are generated or collected in the course of public service delivery, many countries including Australia support public access to and reuse of government data. The provision of data in freely available and reusable formats (so that users can download and interrogate) and under the provision of open licences (without any restriction both in terms of access and fee) is called *open data*.

Due to public demand, some governments mandated their agencies to release data while others left it voluntary (Cerrillo-i-Martínez 2012; Shadbolt et al. 2012). Like other countries Australia publish public datasets in a common site (e.g. www.data.gov.au). These datasets are created, managed, and supplied by different government agencies. That means open data process starts from agency level. There seem to be different practices in the initiative of releasing data between countries, and between agencies within a country. In Australia, there are only 27.4% agencies have adopted 'a strategy' towards open data (McMillan 2013). Similar trend is observed in the USA too; "only 5 of the 169 agencies accounted for 99.37% of all datasets and applications" (Peled 2011, p. 4). That means there could have some distinct characteristics and factors that drive agencies' participation in open data movement. This issue is yet to be investigated. Therefore, this current study aims to explore the antecedents of open data adoption in the context of public agencies. More specifically, the research question for this study is:

*What are the factors that drive (or deter) government agencies to release public data?*

In this study, first, government agency refers to a department of federal or state government responsible for collecting and/or managing data/information as a part of generating or providing service to its people. Then, this study tried to explore the factors that make public agencies to release public data. The decision and process to release data in *open* format is considered as the single most important step to the adoption of open data philosophy because the other activities (e.g. data processing, data reuse etc.) are highly dependent on the released data (Janssen et al. 2012). Doing so, by exploring data from field-study that are also supported by current literature on open data and IS adoption, we developed a research model. It is expected that the findings of the research would reveal some of the major issues pertaining to open data adoption, which would facilitate further provision of open data in the future. As the theoretical contribution, extends organisational behaviour toward technology diffusion.

The remainder of the paper is presented as follows. The next section presents the theoretical background of the current study, followed by discussing the research methodology. Then, the findings of the qualitative field study have been presented while developing the propositions. Followed by a discussion on the field study results, this paper proposes future research directions.

## 2 Background

Everyday government departments collect and produce significant amount of data while they perform their business processes and activities. They keep the data within the organisation and may also make them publicly available – the latter is called *open data* (Zuiderwijk et al. 2014c). *Open data* is considered as a *strategy* of releasing governmental data to anyone at free of cost and without any copyright restrictions (Bertot et al. 2014; Bichard and Knight 2012; Hrynaszkiewicz 2011; Kassen 2013). Open data initiative is considered as one of the most important paradigm-shifts within *open government* movements (Pabón et al. 2013). For political reason, many governments mandated public departments and agencies not only to provide data upon request but release data on the first place without any further copyright obligation to reuse or distribute. Yet, the implementation of open data varies from government to government even from one department to another. Nevertheless, the objective is similar i.e. to enable public access to and reuse of government data reusable formats under open licences.

After examining *open data* policies and their implementation in seven Dutch government departments Zuiderwijk and Janssen (2014) found that the motivation for and capability of opening data varies from departments: some are proactive and highly inspired while others may perform it due to legal obligation. But, successful implementation of open data policy requires effective participation and collaboration between political leaders, public authorities, technologists, and users (Estermann 2014; Zuiderwijk et al. 2014a). Overall, theoretical development of open data is emerging. Not many studies have developed and tested theories/models which explain the adoption process of open data. Among the theories examined no single model seems to dominate - "only rarely was the same theory used more than twice" (Zuiderwijk et al. 2014b, p. iv). A few studies used theories such as Innovation Diffusion Theory, Technology Acceptance Model, Institutional Theory, Motivation theory, actor-network theory.

Rogers (2003)'s Innovation Diffusion Theory (IDT) is one of the most popular that explains organisation adoption-diffusion of an innovation. In open data Estermann (2014) claimed that he used IDT and "create an instrument that allowed measuring the level of adoption of open data". From a (pilot) survey, conducted among 72 respondents, he explored the risks and opportunities, and expected costs and benefits of *open data* but the instrument to measure the level of adoption is somewhat missing. Moreover, he discussed the results in the light of IDT, yet, how he reached to the decisions is unclear. Moreover and most importantly, his study used the five generic innovation-diffusion characteristics as it is, without contextualising; also failed to provide a relative weight of the characteristics (which one is more serious for open data adoption: *compatibility* or *complexity*, for example). At the same time, in order to developed a behavioural model that examines future usage behaviour of open data adopters, Charalabidis et al. (2014) applied the variables from other two highly-used IS theories - Technology Acceptance Model (TAM) and IS Success Model. Their variables are mostly related to the antecedents to open data value generation in e-services.

Janssen et al. (2012) mentioned that the behaviour of many organisations can be explained by Institutional Theory i.e. open data would develop valuable insights among the people who can actually challenge the decision-makers. Zuiderwijk and Janssen (2014) suggested that some organisations have a tendency to publicise their data on similar types of websites and in similar ways – *mimetic isomorphism*. Similarly, by applying Porter's competitive forces model (Hielkema and Hongisto 2013) found that the departments use open data initiative as a competitive advantage.

There are a number of studies that ascertained the drivers and impediments or challenges of open data implementation, discretely. Among them Janssen et al. (2012 presented one of the most comprehensive understanding on barriers of open data adoption while, from prior literature, Zuiderwijk et al. (2012) explored 106 social and technical impediments. Similarly, Barry and Bannister (2014) proposed 20 barriers under six headings: economic, technical, cultural, legal, administrative, risk related. Also, Zotti and La Mantia (2014) identified 4V issues: Volume (the large amount of data), Velocity (the speed of new data arriving), Variety (data with different variety of data), and Veracity (trustworthiness of data). On the other hand, the driving factors for open data movements are yet to be finalised. Prior studies claim that pressure from external and internal environment, economic prospect of agencies as well as of the society as a whole, and technological advancement are the main driving factors for public access on government data (Janssen 2012). Yet, these factors are mostly envisaged from conceptual studies or past literature without much support from empirical evidence, and thereby failed to develop a theory or model. Applying the concepts from existing popular models explaining technology/innovation diffusion in an empirical setting might assist us to understand the adoption factors of open data.

## 3 Research Method

Driven by the objective and nature of the study, field study based qualitative research approach has been adopted for this current study because it is "particularly well suited to new research areas or research areas for which existing theory seems inadequate" (Eisenhardt 1989, pp. 548-549). In other words, little empirical research has been found on the adoption of open data in organisational setting, therefore, an exploratory case study method is used in this instance. This allows us to collect primary data while getting numerous interpretation of a same concept from different respondents. It also assists researchers to explore (or contextualise) specific issues and their impacts on a behaviour, in more detail. Moreover, adoption research of open data is far from maturity; therefore, to study its adoption, the factors and variables are needed to be borrowed from existing literature that also need to be verified by practitioners. In this process, exploration of new factors is likely and be more valuable.

### 3.1 Sample

This current study was conducted in Australia. Australia is one of the originators of Open Government Data (OGD) movements along with New Zealand, Europe, and North America, and one of the current leading countries (with the United States of America, the Scandinavian countries, and the UK government) in national Open Data activities and initiatives (Bauer and Kaltenböck 2012). Since the first inception of open data programs in 2010, till 2014, Australia has released 2135 datasets that have been created from data obtained from several government agencies. Explicitly, Australia's objective is in the line with open data goals: "providing citizen easy access to public data to use and reuse, under open license" (data.gov.au). Hence, the perspectives of the government agency managers captured in this study can enhance the understanding of the overall situation of open data and would provide guideline for organisations within Australia and abroad.

### 3.2 Data collection

For this study, the leading author obtained qualitative data from in-depth interviews conducted with open data practitioners working in government agencies in Australia. Seven interviews with eleven individual respondents from six agencies were undertaken. Among the six agencies three operate at state level, two in federal level, and the rest at local council level. The federal agencies and one state agency have been publishing data under open access policy whereas the rest have not started releasing data yet as and cited resources in regard to remove identifiable entities from raw data as the major issue. Along with data.gov.au the agencies release data through Facebook, Twitter, and YouTube (although all are not reusable). Participants ranged in their position's responsibility including information technologists, data analyst/scientists, open data project managers, and policy analysts.

The participants were selected using convenient sampling, using personal and professional networks. They were interviewed between May and July 2014. In order to ensure consistency, respondents were asked similar questions to evaluate their opinion and experience related to open data adoption in their organisation. They were asked and appreciated to mention any related examples from other organisations too. The interviews were semi-structured, commencing with an open question on perceived motivating and inhibiting factors of open data adoption in public organisations. The average duration of each interview was around 30-40 minutes; each interview session was recorded and later transcribed for analysis.

### 3.3 Data analysis technique

Content analysis (Chan and Ngai 2007) as well as thematic analysis techniques were used for identifying commonly recurring themes (Vaismoradi et al. 2013). During the analysis we applied both inductive and deductive approaches. Inductive analysis was used while developing themes directly from the case data and where previous study is limited. For example, emergence of digital technologies made the open data concept/philosophy as a practical process; this theme could not be related to prior studies. On the contrary, deductive approach was applied to compare and contrast the same themes with from different settings. For example, *institutional pressure* is a well-studied construct in IS literature but its nature and influence is different in the current context, which will be discussed in section 4.2. In brief, the identified factors from the qualitative study were confirmed with existing literature, where possible.

The interviews were transcribed and coded. The coding process involved identifying and arranging concepts in similar groups. This process was double-checked to ensure none of the important themes were missed. Then, the relevant concepts were grouped applying hierarchical method. During this process any discrepancy was settled using support from literature. For example, respondents repeatedly mentioned different sets of resources (such as *financial, human, technical, technological*) although most of them agreed that mere resources do not guarantee the adoption but requires simultaneous strong managerial support. Combining these two themes, prior studies (e.g. Iacovou et al. 1995) suggest *organisational readiness* as a better construct – we espoused the same. As a tool, NVivo 10 was utilised to capture, code, analyse the interviews, and report the findings of the study.

## 4  Findings

The findings of the field study and the associated links with theory are presented in the following section.

### 4.1 Political leadership

Most of the respondents of the field study agreed that although the concept of *open data* has been discussed for a long time within discrete communities, political leaders and their power worked as the most significant driver to the implementation of open data; however, one respondent contended that it is a product of technological advancements including Web2.0. Peled (2011) observed that the mismanagement of shelter and hospital services at the event of Hurricane Katrina in the US in 2005 was mainly due to information-sharing problem among the agency workers. They directed survivors to the already crowded hospitals, and also delayed the evacuation. Hurricane Katrina inspired President Obama to direct agencies to publish all non-classified datasets on the Web. This marks the beginning of open data movement. Sooner, other leaders from UK, Australia, Singapore, Denmark, and Spain joined the movement. Hence, it is the political leaders who institutionalised the concept of open data - their 'political movement' intended to ensure transparency and participation of citizens in governance (Janssen 2012).

Worldwide, political leaders take several initiatives: making public agencies to adopt proper measures to open data, championing open data policy and preparing strategy and directive for departments to act on them, carrying most of the costs for publishing data online (e.g. government of many counties including the US, Denmark, UK, Spain, and Australian carries most of the costs (such as developing and maintaining infrastructure) for publishing data online) (Huijboom and Van den Broek 2011), and preparing an open business environment where private firms and entrepreneurs can participate.

> "Without being started from a policy level and patronised by the government, in terms of both financial and infrastructural support, it is not possible to disseminate open data possibilities". "... You can dream about open data be in cottage industry [by developing and commercialising apps by entrepreneurial initiatives], but to take you there it has to start from government and industry level (Respondent C and D, respectively)

Therefore, the field study suggests that the counties with proactive political leaders who are in favour of public participation through ICT will be well advanced in open data movement. In fact, a clear difference is observed – the countries moved ahead to opening data where the leaders are its supporters or promoters to open data and/or open government compared to countries with less political leadership in such. Therefore, the first proposition is:

> **Proposition 1:** Political leadership will have positive influence on organisational adoption of open data

### 4.2 Institutional pressure

As discussed earlier, open data initiatives are mostly driven by the compelling policies of political governments that organisations cannot avoid. "*Several governments [in the world]* *decided that data related to energy, health, environment, and utility should be available to public so that they can reuse data by developing innovative applications*" (respondent A). Sayogo et al. (2014) also found that, recently, governments mandated several departments and associated private firms to disclose information and to adopt disclosure techniques where visibility, trackability, or traceability is important (such as food/meat supply chain). Keeping aside the comments from respondent D ("open data is a fad" which would take reasonable time to prove), many agencies consume pressure to release data and participate in governments' open data commitments. Such pressure is further intensified by competition between departments. Departments and agencies regularly battle over "political power and institutional legitimacy" and also for "economic fitness" (DiMaggio and Powell 1991, p. 66) – valuable data generated by each agency (e.g. FBI, NASA) are used as a competitive advantage and 'bargaining tool'. In order to survive or do better, agencies realise that open data may encourage different usage of the data than it was initially thought, which consequently inspire innovative products and services (Cerrillo-i-Martínez 2012). Furthermore, government agencies sometimes adopt open data projects because disclosing data is a precondition to enter and to trade into certain markets (such as USA, Japan, EU or Korea) for some products (e.g. coffee, and meat and livestock). It is believed that open data in these industries increases 'product transparency' while assists the markets for 'product differentiation'. Hence, opening data serves as a competitive tool to survive in fierce competition not only for private firms but for a nation as a whole (Sayogo et al. 2014). Studies claim that organisational behaviour is the outcome of institutional pressure, here we espouse the same: public departments would adopt open data policy because of the institutional pressure exerted to them which may come in any form including regulative/coercive, competitive or mimetic pressure, or more in similar (DiMaggio and Powell 1991). Therefore,

> **Proposition 2:** Institutional pressure will have positive influence on organisational adoption of open data

### 4.3 Emergence of technologies in digital market

In last decade the world has experienced a number of disruption and development in digital market especially in computing, telecommunication networks; and the availability, usability, and cost of SmartPhones. Supporting Laudon and Laudon (2004)'s argument, the respondents A, B, and E mentioned that there is a reciprocal relationship between environment and a(n government) organisation: *"more often, government and public departments apprehend or align policies either to respond to the environmental change or to make environmental developments be effectively used by the citizens"*.

Huijboom and Van den Broek (2011) found that technology improvement and technology trend (e.g. mobile apps) drive firms to bring up services that integrate open data. Recent technological developments (e.g. SmartPhone, Internet, Web 2.0) has created an unprecedented explosion of people's computing facility and skills in terms of accessing, storing, manipulating, analysing, linking and distributing data and information (Boulton 2014; Boulton et al. 2011; Rohunen et al. 2014). Moreover, the rapid growth of mobile network is followed by the rise of social networks and with a variety of apps on the mobile device (Huijboom and Van den Broek 2011). Perkmann and Schildt (2015) claimed that open data adoption has been spurred by increasingly widespread use of computers and databases. Similarly, the field study confirmed that such technologies inspired public agencies to consider them for improved service delivery and organisational performance.

> *Access to and use of public data without individual and customisable [electronic] device is expensive, effort-driven, inefficient and thus of no consideration. Just imagine the analogy to PowerPoint slide vs. printed hand-notes – you can reuse the former with least effort* (Respondent B)

> *Internet is now [accessible from] everywhere [using mobile networks]. So, now it is easy for us (i.e. public agencies) if the train schedule is disrupted or there is an un/expected outage…* (Respondent A)

The political leaders' commitment to disclose un-classified public data has been made possible because of the recent developments in technologies that has tremendously reduced the effort to release and reuse open data. Hence, the following proposition is deduced:

> **Proposition 3:** The emergence of digital technologies will have a positive influence on organisational adoption of open data

## 4.4 Interoperability of datasets

As public departments may work at local, national, or regional level data would be generated at different levels. Interoperability of data would assists in the use of data between systems of different agencies, and between agency systems and user's system (Guijarro 2007). Interoperability is often considered as a technical characteristic in IS. In open data context *interoperability* provides the *basic* (technical) specifications that all agencies should apply while preparing and releasing data (Tripathi et al. 2013).

An effective open data initiative ideally is a combined effort from local, national, regional, and global agencies. The field study as well as existing literature identified that it is quite frequent to find "powerful and reach" national metadata but local data are less well-described and suffer from lack of interoperability (Shadbolt et al. 2012). Moreover,

> *… [O]pen data available in heterogeneous and inconsistent formats [that] possesses limited usefulness, in terms of use and reuse* (respondent G)

> *We employed variety of technical systems … Some departments developed new tools or bought systems [that can release data automatically in various forms] … [others] have updated existing systems. [Yet] it is not always the case that we follow a common rule … [consequently] interoperability of data seems a major concern as it continues* (respondent F)

Many studies (e.g. Berners-Lee 2006) believe that developing and adopting common/open standards can resolve interoperability issue. However, agreed with Guijarro (2007) respondent B argued that standard for open data is useful but is not enough and it requires *"a true seamless service delivery to citizens and businesses"*. Other respondents added that, making open data interoperable is also an organisational issue, because *"[departments] need to have available skilled personnel to making data interoperable"*. In spite of the discord solving interoperability issue, every respondent agree that lack of interoperability is a serious barrier to open data adoption. In other words, *perceived interoperability* of open data influence organisations to adopt open data policies (Huijboom and Van den Broek 2011; Janssen et al. 2012). Therefore, the fourth proposition is:

> **Proposition 4:** Perceived interoperability of open data will have positive influence on organisational adoption of open data

## 4.5 Organisational readiness

Although open data movements are mostly driven by political leaders' statements (governments mandate agencies to release certain (number of) datasets), the decision and the process of data release are dependent upon the preparedness of the associated department. Consequently, the success of open data adoption varies among departments within a same government.

O*rganisational readiness* is defined as "the availability of the needed organisational resources for adoption" (Iacovou et al. 1995, p. 467). It evaluates whether a firm has sufficient preparedness in terms of IS sophistication (both in terms of technology and users) and economic costs. Prior studies agree that IS innovations such *open data* requires an extensive resources that includes financial, technical, and skilled human resources (Ramdani et al. 2009). Financial resources express an organisation's capital availability to invest in an IS innovation. More often, organisations suffer from limited financial resources for initiating and continuing open data projects. Additionally, selling data is a good source of revenue for many agencies/departments, which makes them reluctant to release data. In other words, *perceived loss of revenue* or *extra income* is an economic barrier to open data adoption that an organisation considers seriously. *"We need to find out a new business [and] revenue model that should be sustainable for next couple of years at least"* – respondent H. In the positive side, data release through *open data* portals is cheaper than rendering them into reports and applications and thus saves a lot for many departments (Hendler et al. 2012). Moreover, reuse of open data in the downstream promises a lot of economic benefits (Conradie and Choenni 2014). Hence, organisations need to be confident that open data projects would be economically justifiable. Apart from financial preparedness, open data projects require skilled and technical human resources, and various IT hardware and software (Ramdani et al. 2009).

> *Lack of technology surely hinders open data growth [in organisations]; however, lack of skilled and experienced human resources is even worse because you already [have] invested and getting low ROI than it should be* (respondent B)

> *In fact, a few departments bought expensive systems but do not know how to use* (respondent I)

Extant literature as well as the respondents of the field study highlighted that organisational readiness is important for open data adoption. Therefore, the next proposition is:

> **Proposition 5:** Organisational readiness will have positive influence on organisational adoption of open data

## 4.6 Management commitment

Prior studies recognise the organisational commitment toward open data as an important organisational variable (Linders 2013). Respondents agreed that open data initiatives would not be successful if there is no support from top management and is not embedded into the organisation's mission. Accordingly, management must develop and implement an effective strategy by assigning appropriate level of authority. For instance, releasing data and information about heritage collections is a fundamental commitment of heritage organisations that makes least barriers to open data (Estermann 2014). Also, *"[operational] managers usually enjoy the decisive power to captive data, and release upon [huge] request than releasing data in the first place"*. Respondents A, B, and D emphasised on the support from senior management for an open access policy. In that case, management must realise that opening data would increase goodwill and recognition while it also increases user satisfaction (through fulfil the institute's mission), for instance. Furthermore, a vital institutional concern is *perceived loss of control over data*. Each agency and department often competes with one another over resource, recognition, influence and control, and autonomy. In such competition datasets are considered as a valuable asset. Hence, some agencies would be reluctant to adopt open data. Respondent C claims that top management's position toward open data dynamism is highly related with *awareness* and *knowledge* of the associated managers. *"Many department like us lag in open data initiatives because few of our top $#*! [managers] have no #$*@!$ idea about the prospect and possibility [of open data]... They are highly conservative, committed to only the rich, [and] believe on costs but not on benefits"* – respondent E.

A strong influence of the commitment of organisation's management toward open data movement, therefore the final proposition is:

> **Proposition 6:** organisation's management commitment will have positive influence on organisational adoption of open data

Combining the developed propositions, the research model is presented in next page (Figure 1).

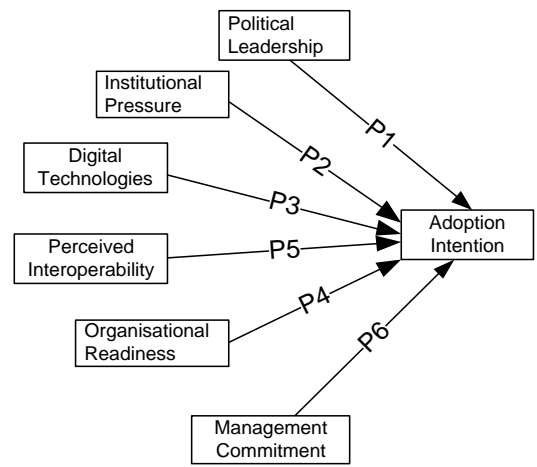

*Figure 1: The research model explaining open data adoption in government organisations*

## 5 Discussion

Unstructured data are less useful as crude oil; once processed, data can power a nation like refined oil (Palmer 2006). Hence, data generated and stored by government agencies are valuable for a nation but only when these are translated to structured data. To allow the use of public data at broader level is to open them to public access provided no legal or security/privacy issue. Therefore, social and technology scientists advocate that government agencies should ease the process to publicize public data so that data can be utilised with ease. In this process, the respondents of this study identified that first of all *management commitment* is vital to ensure an agency to adopt proactive publication stance. Also, the department or agency needs to be aware of the value proposition and should understand the culture of openness in regard to open access and proactive publication (opposite to release data upon request). Accordingly, the management would undertake cost benefit analysis for the open data projects. This study found that many agencies were willing to join *the next big thing* but were less active because of the resource constraints. Yet, the relationship between the departmental size and financial support for open data initiative is mixed. Some respondents believe that large agencies are in a better position in terms of resources. These departments have been operating for a long period of time and therefore have developed their expertise and have resources in place. Moreover, their availability of and access to financial and expertise and thus to technologies are comparatively better than smaller agencies. On the other hand, the rest believe that large agencies often have large and more complex datasets that are challenging to manage. The process of data preparation therefore is complex, time-consuming, and requires specialised personnel. This is more so if they have offices in multiple locations (in different States/territories) (McMillan 2013). Additionally, large agencies usually are already overwhelmed with daily tasks and thus are less innovative and agile in their approach to new technologies. Future studies are required to better understand the relationships between the *firm size* and open data adoption.

Interestingly, some (senior and influential) managers consider releasing data as a part of voluntary social responsibility, with low priority. Sometimes their internal evaluations ascertain little prospect of public data use. But, data scientists and advocates suggest that data should be released on the first place; data use will be a natural process, which may take time. This approach also solves the *chicken and egg* problem of data release and data use. Hence, in order to ensure access of public to their data, political leaders can play a vital role. If it is too early to mandate, government need to set defined guidelines that include timing and volume/quality of dataset release, respondents believe. More importantly, independent watchdog with authority should monitor whether the agencies follow the guidelines accordingly, and investigate further scope of data release after having discussion with different stakeholders such as community representatives, technologists, (IT) business entrepreneurs, citizens (as both source and users of data), researchers, and agency managers.

This study further confirms that data interoperability is critical for open data adoption; data from different sources should be ready for further use with minimal effort. Where possible, data should be provided in multiple formats to support different user groups. Such capability would allow firms of *novel use* of existing data that is beyond the original objective or capability of many agencies. Although it is a technological issue this study points out that the decision of providing interoperable data is an interdepartmental managerial decision, which involves effort in terms of budget and time. Meanwhile, the development of various digital devices speeds up data use. Many software/applications increasingly are available for different platforms (e.g. for android, windows, or mac operating systems). Also, mobile devices now can collect, read, and process data in different formats. Further development in digital technologies would enhance open data use in future. All together, a collaborative effort from technologists, business analysts, data scientists, and agency managers would be useful for open data movement.

## 6　Conclusion and Future Research Direction

Many governments have been promising public to be open and transparent through citizens' unrestricted access to and use of public data; however, such commitments are very difficult to keep if the associated factors are not known. For example, it is found that, in order to comply with the mandate from federal government, several agencies only publish data that are less importance.  Using a qualitative approach, this study develops an initial research model that explains the influence of six factors on open data adoption.

The research model presents the antecedent factors influencing organisational adoption of open data — a newly emerging phenomenon that has limited empirical evidence. Future research will have the opportunity to further test the model and explore some constructs and their relationships (e.g. *size* of the agency and *resource readiness*). In order to get more insight, the model could be tested with longitudinal data that compares the difference in perception of the organisations before and after adopting open data policies. From only seven interviews, we realised that a comprehensive view might not be obtained. Insights into the various government departments across Australia suggest that such characteristics seem to be similar. In future, experience from organisations involved in various stages of technology adoption, (continued) use, and routinisation and assimilation with existing business activities might reveal the nature of open data diffusion and therefore would be worthwhile to study. Moreover, separate models could be developed for different departments that differ in service generation and value creation to the society.

## 7　Reference

Barry, E., and Bannister, F. 2014. "Barriers to Open Data Release: A View from the Top," *Information Polity* (19:1), pp. 129-152.

Bauer, F., and Kaltenböck, M. 2012. *Linked Open Data: The Essentials a Quick Start Guide for Decision Makers*. Vienna, Austria: edition mono/monochrom.

Berners-Lee, T. 2006. "Linked Data-Design Issues." from http://www.w3.org/DesignIssues/LinkedData.html

Bertot, J. C., Gorham, U., Jaeger, P. T., Sarin, L. C., and Choi, H. 2014. "Big Data, Open Government and E-Government: Issues, Policies and Recommendations," *Information Polity* (19:1), pp. 5-16.

Bichard, J.-A., and Knight, G. 2012. "Improving Public Services through Open Data: Public Toilets," *Proceedings of the ICE-Municipal Engineer* (165:3), pp. 157-165.

Boulton, G. 2014. "The Open Data Imperative," *Insights: the UKSG journal* (27:2), pp. 133-138.

Boulton, G., Rawlins, M., Vallance, P., and Walport, M. 2011. "Science as a Public Enterprise: The Case for Open Data," *The Lancet* (377:9778), pp. 1633-1635.

Cerrillo-i-Martínez, A. 2012. "Fundamental Interests and Open Data for Re-Use," *International Journal of Law and Information Technology* (20:3), pp. 203-222.

Chan, S. C., and Ngai, E. W. 2007. "A Qualitative Study of Information Technology Adoption: How Ten Organizations Adopted Web-Based Training," *Information Systems Journal* (17:3), pp. 289-315.

Charalabidis, Y., Loukis, E., and Alexopoulos, C. 2014. "Evaluating Second Generation Open Government Data Infrastructures Using Value Models," in: *Proceedings of the 2014 47th Hawaii International Conference on System Sciences*. IEEE Computer Society, pp. 2114-2126.

Conradie, P., and Choenni, S. 2014. "On the Barriers for Local Government Releasing Open Data," *Government Information Quarterly*), pp. S10-S17.

DiMaggio, P., and Powell, W. 1991. "The Iron Cage Revisited: Institutional Isomorphism and Collective Rationality in Organizational Fields," in *The New Institutionalism in Organizational Analysis,* W. Powell and PJ (eds.). pp. 63-82.

Eisenhardt, K. M. 1989. "Building Theories from Case Study Research," *The Academy of Management Review* (14:4), pp. 532-550.

Estermann, B. 2014. "Diffusion of Open Data and Crowdsourcing among Heritage Institutions: Results of a Pilot Survey in Switzerland," *Journal of theoretical and applied electronic commerce research* (9:3), pp. 15-31.

Guijarro, L. 2007. "Interoperability Frameworks and Enterprise Architectures in E-Government Initiatives in Europe and the United States," *Government Information Quarterly* (24:1), pp. 89-101.

Hendler, J., Holm, J., Musialek, C., and Thomas, G. 2012. "Us Government Linked Open Data: Semantic. Data. Gov," *IEEE Intelligent Systems* (27:3), pp. 25-31.

Hielkema, H., and Hongisto, P. 2013. "Developing the Helsinki Smart City: The Role of Competitions for Open Data Applications," *Journal of the Knowledge Economy* (4:2), pp. 190-204.

Hilvert, J. 2013. "Slow Progress on Government's Open Data Effort."　Retrieved 04/08, 2015,　from　http://www.itnews.com.au/News/334212,slow-progress-on-governments-open-data-effort.aspx

Hrynaszkiewicz, I. 2011. "The Need and Drive for Open Data in Biomedical Publishing," *Serials: The Journal for the Serials Community* (24:1), pp. 31-37.

Huijboom, N., and Van den Broek, T. 2011. "Open Data: An International Comparison of Strategies," *European journal of ePractice* (12:1), pp. 1-13.

Iacovou, C. L., Benbasat, I., and Dexter, A. S. 1995. "Electronic Data Interchange and Small Organizations: Adoption and Impact of Technology," *MIS Quarterly* (19:4), pp. 465-485.

Janssen, K. 2012. "Open Government Data and the Right to Information: Opportunities and Obstacles," *The Journal of Community Informatics* (8:2).

Janssen, M., Charalabidis, Y., and Zuiderwijk, A. 2012. "Benefits, Adoption Barriers and Myths of Open Data and Open Government," *Information Systems Management* (29:4), pp. 258-268.

Kassen, M. 2013. "A Promising Phenomenon of Open Data: A Case Study of the Chicago Open Data Project," *Government Information Quarterly* (30:4), pp. 508-513.

Laudon, K. C., and Laudon, J. P. 2004. "Management Information Systems: Managing the Digital Firm," *New Jersey* (8).

Linders, D. 2013. "Towards Open Development: Leveraging Open Data to Improve the Planning and Coordination of International Aid," *Government Information Quarterly* (30:4), pp. 426-434.

MacGunigal, M. 2014. "Regrets About Open Data? Only That We Didn't Start Sooner," in: *The Public Manager*. Association for Public Development (ASTD).

Mazumder, S. 2014. "Turning up the Heat on Open Data," in: *IBM Data Magazine*. IBM Data Magazine.

McMillan, J. 2013. "Open Public Sector Information: From Principles to Practice," *Australian Government-Office of the Australian Information Commissioner*).

Pabón, G., Gutiérrez, C., Fernández, J. D., and Martínez-Prieto, M. A. 2013. "Linked Open Data Technologies for Publication of Census Microdata," *Journal of the American Society for Information Science and Technology* (64:9), pp. 1802–1814.

Peled, A. 2011. "When Transparency and Collaboration Collide: The USA Open Data Program," *Journal of the American society for information science and technology* (62:11), pp. 2085-2094.

Perkmann, M., and Schildt, H. 2015. "Open Data Partnerships between Firms and Universities: The Role of Boundary Organizations," *Research Policy* (44:5), pp. 1133-1143.

Ramdani, B., Kawalek, P., and Lorenzo, O. 2009. "Predicting Smes' Adoption of Enterprise Systems," *Journal of Enterprise Information Management* (22:1/2), pp. 10-24.

Rogers, E. M. 2003. *Diffusion of Innovations*, (5Th ed.). New York: Free Press

Rohunen, A., Markkula, J., Heikkila, M., and Heikkila, J. 2014. "Open Traffic Data for Future Service Innovation: Addressing the Privacy Challenges of Driving Data," *Journal of Theoretical and Applied Electronic Commerce Research* (9:3), pp. 71-89.

Sayogo, D. S., Zhang, J., Pardo, T. A., Tayi, G. K., Hrdinova, J., Andersen, D. F., and Luna-Reyes, L. F. 2014. "Going Beyond Open Data: Challenges and Motivations for Smart Disclosure in Ethical Consumption," *Journal of Theoretical and Applied Electronic Commerce Research* (9:2), pp. 1-16.

Shadbolt, N., O'Hara, K., Berners-Lee, T., Gibbins, N., Glaser, H., and Hall, W. 2012. "Linked Open Government Data: Lessons from Data. Gov. Uk," *IEEE Intelligent Systems* (27:3), pp. 16-24.

Tripathi, R., Gupta, M., and Bhattacharya, J. 2013. "Effect of Organizational Factors on Interoperability Adoption for Indian Portals," *Transforming Government: People, Process and Policy* (7:3), pp. 285-308.

Vaismoradi, M., Turunen, H., and Bondas, T. 2013. "Content Analysis and Thematic Analysis: Implications for Conducting a Qualitative Descriptive Study," *Nursing & health sciences* (15:3), pp. 398-405.

Zotti, M., and La Mantia, C. 2014. "Open Data from Earth Observation: From Big Data to Linked Open Data, through Inspire," *Journal of e-Learning and Knowledge Society* (10:2), pp. 91-100.

Zuiderwijk, A., Gaseó, M., Parycek, P., and Janssen, M. 2014a. "Special Issue on Transparency and Open Data Policies: Guest Editors' Introduction," *Journal of Theoretical and Applied Electronic Commerce Research* (9:3), pp. I-IX.

Zuiderwijk, A., Helbig, N., Gil-García, J. R., and Janssen, M. 2014b. "Special Issue on Innovation through Open Data: Guest Editors' Introduction," *Journal of theoretical and applied electronic commerce research* (9:2), pp. i-xiii.

Zuiderwijk, A., and Janssen, M. 2014. "Open Data Policies, Their Implementation and Impact: A Framework for Comparison," *Government Information Quarterly* (31:1), pp. 17-29.

Zuiderwijk, A., Janssen, M., Choenni, S., and Meijer, R. 2014c. "Design Principles for Improving the Process of Publishing Open Data," *Transforming Government: People, Process and Policy* (8:2), pp. 185-204.

Zuiderwijk, A., Janssen, M., Choenni, S., Meijer, R., and Sheikh_Alibaks, R. 2012. "Socio-Technical Impediments of Open Data," *Electronic Journal of e-Government* (10:2), pp. 156-172.